\documentstyle[12pt,titlepage]{article}
\newcommand{\bq}{\begin{equation}}
\newcommand{\eq}{\end{equation}}
\newcommand{\ang}[1]{$#1^\circ$}         
\begin{document}
\begin{titlepage}
 \begin{center}
{\large\bf
 Multiple Jumps and Vacancy Diffusion \\
in a Face-Centered Cubic Metal\\}
\vspace{1.0cm}
{\sc G.\ De Lorenzi \rm (*)(***) and \sc F.\ Ercolessi \rm (**)(***) \\}
 \end{center}
\vspace{0.4cm}
{(*) \it Centro di Fisica del Consiglio Nazionale delle Ricerche,
      I-38050 Povo, Trento, Italy\\}
{(**) \it International School for Advanced Studies,
     I-34014 Trieste, Italy\\}
{(***) \it Materials Research Laboratory,
     University of Illinois at Urbana-Champaign,
     Urbana, IL 61801, USA\\}
\vspace{1.0cm}

\noindent
{\small
}
\vspace{3.0cm}

{\small
cond-mat/9208008 - To appear on {\sl Europhysics Letters}.
}

\clearpage
{\bf Abstract.}
The diffusion of monovacancies in gold has been studied by computer
simulation. Multiple jumps have been found to play a central role
in the atomic dynamics
at high temperature, and have been shown to be responsible
for an upward curvature in the Arrhenius plot of
the diffusion coefficient.
Appropriate saddle points on the potential energy
surface have been found, supporting the interpretation of
vacancy multiple jumps as  distinct migration mechanisms.

\end{titlepage}

Several face-centered cubic metals,
such as Ag, Cu, Au, Ni, and Pt, exhibit an upward curvature
in the Arrhenius plot of the atomic self-diffusion coefficient
close to the melting point \cite{exp.gold}.
To explain this effect, three mechanisms have been proposed \cite{debate}:
(a) different types of defects (such as
    monovacancies and divacancies) contributing to the diffusion process
    with different activation energies \cite{peterson};
(b) an intrinsic temperature dependence of enthalpy and entropy changes
    associated with the formation and migration of monovacancies
    \cite{Gilder};
(c) contributions from double jumps of monovacancies
    \cite{DaFano}.
 In this Letter we report molecular
dynamics simulations in gold aimed at characterizing qualitatively and
quantitatively the effect of correlated multiple jumps on the temperature
dependence of the diffusion coefficient. Furthermore,
from detailed studies of the
potential energy surface, we provide a new microscopic interpretation of---at
least---the double jump mechanism
\footnote{It should be pointed out that our simulations do
not address mechanisms (a) and (b) above, which therefore  remain as further
possible factors contributing to the curvature
as suggested by some experiments.}.

It has already been known for some time
that correlated jumps  occur in the diffusion of vacancies in solids
close to melting \cite{DaFano,Bennett,FJ}, but no statistically
relevant measurements of this effect have been reported yet.
Our results, based on the analysis of a large number of events, show that
multiple jumps are present and very effective
in bending the Arrhenius plot near
the melting point. An interpretation of their microscopic nature has also been
needed, in order to integrate them into the framework  of those recent
developments in the theory of diffusion
that explicitly take into account dynamical effects \cite{Toller}.

Empirical or semi-empirical many-body Hamiltonians
developed in the last years have proved to be
much more realistic for metals than pair potentials used in the past,
in particular to describe surfaces or defects.
The system we have used in our simulation consists of 255 particles with
periodic boundary conditions, interacting via the ``glue'' many-body
potential \cite{PM}, which is well suited for our study since
it reproduces rather well the melting temperature
and the thermal expansion of Au,
as well as the $T=0$ energetics of a monovacancy.
We have analyzed all the vacancy jumps in runs
  performed at twelve different temperatures, between 1000K and 1550K,
  at steps of 50K.
All the runs are done in the microcanonical ensemble, where
temperature is defined by time averaging the total kinetic energy.
At each temperature, the lattice spacing is obtained
by an independent zero-pressure simulation of a perfect crystal.
Since melting occurs near 1350K in this model \cite{PM},
runs at 1400K, 1450K, 1500K and 1550K refer to an overheated crystal
which, in the absence of free surfaces,
remains stable within the simulation time scale
(mechanical instability was reached around 1600K).
We find these runs to be very useful to observe the trends
of the diffusion mechanisms at high temperatures,
even if they cannot be directly compared with experimental results.
With the exception of the runs at 1000K, 1050K and 1100K,
where the jump rate is particularly low,
the runs are made long enough to contain $\sim 800$ jumps
in order to obtain sufficient statistics of rare
correlated jumps.
This requires a run length of 57 ns ($8\times 10^6$ MD steps)
at $T=1150$K, and progressively shorter runs at higher temperatures.

{}From the  Arrhenius plot shown in
fig.~\ref{fig:difcoe} (filled circles), it can already
be seen that migration of {\it monovacancies\/} alone---the
only type of defect present in our
simulation---produce a measurable curvature. The curve bends just below $T_m$
and continues with a higher slope into the superheated region.
The quantity plotted is the atomic migration contribution to the
diffusion coefficient, defined as
   \bq
   D = \frac {\langle u^2 (t) \rangle - \langle u^2_{\rm bulk} \rangle }{6t},
   \label{eq:msd}\eq
   where $t$ is the run length, $u^2 (t) = |{\bf r}(t) - {\bf r}(0)|^2$,
         $u^2_{\rm bulk}$ is the asymptotic value of the same
         quantity for a perfect bulk (not diffusing),
   independently calculated and subtracted in order to retain only
   the diffusive part,
   and the angular brackets indicate average over all the atoms.
The atomic diffusion coefficient, as usually defined, is $D$ multiplied
by the ratio of the concentration of vacancies in the system
at the given temperature
to the concentration in our model system
(fixed to be 1:256 at all temperatures).

To determine the
 presence and role of vacancy multiple jumps, we have analyzed
 all the jump events with respect to both their
 relative times of occurrence and of the atomic displacements involved.
In a {\it vacancy multiple jump\/} two or more atoms move simultaneously.
Each atom moves by only one lattice spacing, $d$,
but the vacancy, as a result of this
concerted motion, is displaced by one or more lattice spacings, depending
on the number of atoms involved and their relative directions.
In a double jump, for example,
if the two atoms move in the same direction the vacancy is
displaced by $2d$ , but if the two atoms move at an angle of \ang{120}
the vacancy will be displaced by only $d$.

     We have first analyzed the jump events in terms of their relative
     times of occurrence.
     At each time step, the occupation number of each Wigner-Seitz
     cell in the fcc lattice is computed. Every change in the
     occupation numbers
   (``event'') is recorded, along with the time step at which it occurs. A
    large fraction of the jumps has been found
     to be followed almost
     immediately [i.e., within a Debye period,
     corresponding to about 40 MD steps (0.29 ps)] by a return jump into
     the original position.
     It is very likely that these event pairs are not real jumps,
     but rather occasional excursions of a neighbouring atom into
     the cell containing the vacancy.
     Such jumps are considered as miscounts and eliminated from the analysis.
     If we consider all the jumps followed by a return
     jump within a time $\tau$ and calculate the average delay $\delta$,
     we observe that when $\tau$ becomes larger than
     about 50 steps (0.36 ps), $\delta$ tends to stabilize around
     25 steps, nearly independently of temperature.
     We have therefore taken $\tau_R=25$ steps (0.18 ps) as the average delay
     of immediate returns, and removed all the
     jumps followed by a return within $3\tau_R=75$ steps.
     Assuming an exponential distribution, this should
     eliminate about 95\% of these miscounts. After this filtering,
     the distribution of delays between successive jumps
     at $T=1450$K
     is shown in fig.~\ref{fig:delays}.
     For large times the semi-logarithmic plot is linear, as expected for
     uncorrelated jumps.
     For delays of less than 200 steps (1.43 ps), however, strong deviations
     from linearity indicate
      the presence of correlated jumps. The same feature, even
     if not so pronounced, is found also at the lower temperatures, down to
     $1250$K.
     Qualitatively similar results have been found in a
     simulation study of vacancies in Al \cite{DaFano}. Thanks to the
     much larger statistics of events, we can here provide statistically valid
     information about their contribution to the Arrhenius plot, and we also
      provide an interpretation of their microscopic mechanism.

     We have classified all jumps into
     single, double, triple, etc., by grouping together
     those events separated in time by less than $\tau_M=200$ steps.
     The temperature dependence of the frequencies for the different groups
     is reported in fig.~\ref{fig:freqs}, referring to the range
from 1150K to 1450K
[computational limitations prevented us to obtain sufficient statistics
 of multiple jumps for $T<1150$K,
 while at 1500K and 1550K (far into the superheated region)
 correlated jumps become so complex to make this analysis
 impractical to do]. The
     apparent activation energies are not simple multiples of the energy for
    single jumps,
     as would be expected if they were extracted from a distribution
     of uncorrelated jumps. The contribution $D_S$ of single jumps to the
  diffusion coefficient, represented by the open circles in the
  Arrhenius plot of fig.~\ref{fig:difcoe}, even exhibits a small downward
  curvature
  at the highest temperatures (superheated region), in contrast
  with the upward curvature of the total $D$.
  {}From this second step of our analysis, it is clear that
  the upward curvature observed in our simulation is due
  to correlated jumps.

      We have proceeded in analyzing these correlated events in terms of the
     angles between the directions of successive  vacancy jumps.
      Reasonable statistics was available only for double jumps. In the
     fcc lattice there are five  possible angles between the directions
     of two vectors joining three consecutive lattice sites:
     \ang{0} (collinear), \ang{60}, \ang{90}, \ang{120}
     and \ang{180} (return jumps).
     Given a vacancy in a specific lattice site, there are
     $m$ equivalent ways it can perform a given double jump,
     where $m=12$, $48$, $24$, $48$, and $12$ for
     \ang{0}, \ang{60}, \ang{90}, \ang{120} and \ang{180}
     double jumps, respectively. For a series of random jumps,
       the frequency, $\Gamma$, divided by the multiplicity, $m$,
     should then be uniformly distributed among the five angles.
       This is what we find, within the statistical uncertainty,
       for pairs of jumps with delays $\tau >\tau_M$.
       But for $\tau < \tau_M$ the distribution is rather different,
       as shown in fig.~\ref{fig:angles}.
       We can observe some clear trends:
       a small increase above the random ratio for \ang{0},
       a definite increase for \ang{60},
       a small decrease for \ang{90} and
       a definite decrease for \ang{120} double jumps.
       \ang{180} double jumps also increase, but behave quite erratically
        with temperature.
        These last events correspond to return jumps of the same atom
         and do not contribute to the diffusion. In agreement with the
       previously given definition, in the following
       we will call {\it double jumps\/} only those events involving
        two atoms.
       The apparent activation energies for the separate groups
        are: $1.9, 1.6, 1.9$ and $1.6\pm 0.2$ eV
       for \ang{0}, \ang{60}, \ang{90} and \ang{120}
       double jumps, respectively.

       The question now arises if these correlated jumps correspond to
       dynamical effects or to different jump mechanisms \cite{FJ}.
       The first situation would imply
       the fast successive crossing of two single jump barriers
       in series, the latter the crossing of a different, higher
       barrier, specific for each double jump.
       To shed some light on this point, we
       have thoroughly investigated the potential energy surface in
       $3N$-dimensional space \cite {gianni},
       around possible configurations for double jump saddle-points.
       We proceeded by trial and error, first looking for watersheds
       between equilibrium configurations with minimization techniques,
       and then looking for extrema in the intermediate regions, always
      letting all the $3N$ atomic coordinates relax in this process.
       The calculation  has been done at the equilibrium density for $T=1350$K.
       Relatively low energy paths for double jumps can in fact
       be found. The corresponding saddles are not as simple
       as the traditional saddle point for single jumps
       (fig.~\ref{fig:saddlesj}), which is quadratic and has only one
        unstable normal mode,  but are instead combinations of minima, saddle
       points and crossroads (saddle
       points with two unstable normal modes), as shown, e.g. for the case of
       \ang{0} double jumps in fig.~\ref{fig:saddledj}.
       However, the difference
       in energy between these local structures is never larger than 0.02 eV.
       The energies of the lowest points in the barriers for the
       various double jumps with respect to equilibrium are:  1.93, 1.92,
       2.29 and 2.66 eV for \ang{0}, \ang{60}, \ang{90} and \ang{120}
      double jumps, respectively. These values are to be compared with the
      apparent
       activation energies evaluated from MD and reported above.
       Taking into account the errors in
       the MD estimates, and the fact that such a comparison is
       valid only at low $T$, the energies for
       \ang{0}, \ang{60} and \ang{90} jumps can be considered rather close.
        This agreement is an
        indication that double jumps constitute distinct jump mechanisms,
        corresponding to the passage over characteristic energy barriers.
        In fact, we have also verified that the distance between
       the two jumping atoms remains close to the first-neighbour
        distance throughout the jump, in most \ang{0}, \ang{60} and \ang{90}
        jumps.
        The MD estimate of the activation energy for \ang{120}
        jumps is, on the other hand, sensibly lower than
        the height of the barrier found for this mechanism.
        However, the frequency of \ang{120} jumps is
        much                              
        smaller than the other types of double jumps,
        and these events cannot really be distinguished from those
        that would be expected for a distribution of pairs
        of successive uncorrelated single jumps.

        Finally we note that the anomalous number of return jumps
        could be due to the complicated structure
        of the saddles, where trajectories can be momentarily trapped and
        then turn back.
        As shown by fig.~\ref{fig:saddledj}, successful completion
        of a double jump also requires the unstable normal mode
        to change near the saddle point.
        Moreover, we have found a second family of saddle
        points for single jumps on the sides of the
        traditional one (fig.~\ref{fig:saddlesj}).
        In these, the jumping atom is
        still placed halfway between the two half-vacancies, but, instead of
        lying in the (100) plane, it sits
         either below or above it by about 30\% of the
        nearest-neighbour distance. These lateral saddles have the same
         complicated structure
        as the saddles for double jumps (fig.~\ref{fig:saddledj}), with a
        saddle point energy of $1.18$ eV, and a local minimum of $1.15$ eV,
        lower than the energy of the
        traditional saddle point for single jumps, $1.34$ eV.

In summary, the migration of monovacancies in gold, as simulated by
molecular dynamics, exhibits an Arrhenius
plot with a pronounced upward curvature
at high temperature. This curvature is due to multiple jumps,
in which two or more atoms jump simultaneously.
An analysis of the potential energy surface
reveals the presence of distinct saddle points
appropriate for double jumps, and
supports the interpretation of correlated jumps
as distinct migration mechanisms.

\noindent\begin{center} *** \end{center}

Useful comments by Gert Ehrlich, Giulia Galli, Gianni Jacucci,
Enrico Smargiassi and by a referee are gratefully acknowledged.
This work has been performed partly under the
 Progetto Finalizzato CNR ``Sistemi informatici e calcolo
 parallelo''
and under the U.S.\ Department of Energy Grant No. DEFGO2-91ER45439.

\clearpage
\section*{Figure Captions}
\newcounter{fignum}
\begin{list}{Figure\ \thefignum .}{\usecounter{fignum}}

\item
Atomic migration contribution to the tracer diffusion coefficient $D$
via monovacancies.
Filled circles indicate $D$ as
measured directly from Eq.~(\ref{eq:msd}); diamonds indicate
$D_{\rm all}$ as evaluated from the jump frequency $\Gamma$,
with all the jumps treated as if they were single, uncorrelated jumps;
and open circles indicate the contribution $D_S$
from only single jumps.
$d$ is the nearest-neighbour distance and
$f_S=0.78146$ is the geometrical correlation factor for single jumps.
Apparent activation energies are shown.
\label{fig:difcoe}

\item
Time delays between successive jumps of the vacancy.
The bin width is 200 steps.
\label{fig:delays}

\item
Frequencies of single (S), double (D), triple (T) and quadruple (Q) jumps.
Error bars correspond to a standard deviation.
The lines, whose slopes are given as activation energies,
are obtained by a weighted fit.
\label{fig:freqs}

\item
Distribution of angles between the jump directions of pairs of successive
jumps. The jump frequency $\Gamma$ is divided by the geometrical
multiplicity $m$ (see text).
\label{fig:angles}

\item
Saddle on  the $3N$-dimensional potential energy surface for
single jumps. The plot shows a section of the hypersurface along the
directions of
the unstable normal mode (the jumping mode), $\eta$, and the softest
stable normal
mode, $\zeta$. Lengths along  $\eta$ and $\zeta$ are in {\AA} and the
energy is
reported relatively to the energy of the equilibrium configuration.
The atomic
positions in the saddle point configuration and the atomic displacements
corresponding to $\eta$, projected onto the (100) plane, are also shown.
Note that $\eta$ and $\zeta$ are both $3N$-dimensional vectors, but their
components along the cartesian coordinates of the nonjumping atoms are so
small that cannot be detected in the figure.
\label{fig:saddlesj}

\item
Saddle for \ang{0} double jumps.
Conventions are as in fig.~\ref{fig:saddlesj}.
The central configuration is a local minimum.
$\zeta$ is the direction of the softest mode.
Along the second axis, in order to clearly show the relevant structures
of the energy surface, we have chosen two different half-directions
$\eta_-$ and $\eta_+$ for the negative and positive part.
This is why there is a sharp edge where the two sections meet.
The two sketches on the left show the atomic displacements
in the crossroad positions.
Direction $\eta_-$ corresponds to the first half of the jump: the atom
in front is slower than the one behind it, so that their
separation decreases (as also seen by the overlap of the
two jumping atoms).
Direction $\eta_+$ corresponds to the second half of the jump:
the atom in front is faster.
Completion of a double jump therefore requires a change of
the unstable normal mode to occur near the central position.
\label{fig:saddledj}
\end{list}
\end{document}